# Mechanical Properties of the Electric Field: A Novel Prediction Derived from the Field's Mass and Stress

Eliahu Cohen[1], Paz Beniamini[2], Doron Grossman[2],
Lawrence P. Horwitz[1,3,4,5], Avshalom C. Elitzur[5]

[1] School of Physics and Astronomy, Tel-Aviv University, Tel-Aviv 69978, Israel
[2] Department of Physics, The Hebrew University, Jerusalem 91904, Israel
[3] Bar-Ilan University, Department of Physics, Ramat Gan, Israel
[4] College of Judea and Samaria, Ariel, Israel
[5] Iyar, Israeli Institute for Advanced Research, Rehovot, Israel

*An experiment is proposed which can distinguish between two approaches to the reality of the electric field, and whether it has mechanical properties such as mass and stress. A charged pendulum swings within the field of a much larger charge. The two fields manifest the familiar apparent curvature of their field-lines, "bent" so as not to cross each other. If this phenomenon is real, the pendulum's center of mass must be proportionately shifted according to its lines' curvature. This prediction has no precise counterpart in the conventional interpretation, where this curvature is a mere superposition of the two fields' crossing lines. This empirical distinction, meriting test in itself, further bears on several unresolved issues in classical, quantum and relativistic electromagnetism.*

The electromagnetic field is the most prevalent among the four known to physics. As such, even after centuries of intensive research, it keeps offering fresh insights into the nature of physical fields. An especially interesting issue of this kind concerns the electric field-lines: Are they real ingredients of the field, capable of curving as they appear to do under appropriate circumstances? Or is this curvature only a mathematical abstraction not to be taken literally?

Earlier [1] we have taken the former position. Our conclusion was based on some well-established proofs that the electric field carries also some of the charge's mass, and that, when the charge accelerates, its field produces stress force, manifested in the field-lines curvature:

$$F_s = \frac{E^2}{4\pi R_c} \equiv \frac{E^2 a \sin\theta}{4\pi c^2}, \qquad (1)$$

where $R_c$ is the curvature's radius, $E$ and $a$ are the charge's electric field and acceleration respectively, and $\theta$ the angle between the acceleration's and the field-lines' directions [2-3 and references therein].

We further argued [1] that this stress force can be straightforwardly derived from the electromagnetic stress tensor

$$\sigma_{ij} = \frac{1}{4\pi}[E_i E_j + H_i H_j - \frac{1}{2}(E^2 + H^2)\delta_{ij}] \qquad (2)$$

by the relation

$$(F_s)_i = \frac{\partial \sigma_{ij}}{\partial x_j}, \qquad (3)$$

thus incorporating also the magnetic contribution which is an integral part of the electromagnetic tensor $F_{\mu\nu}$.

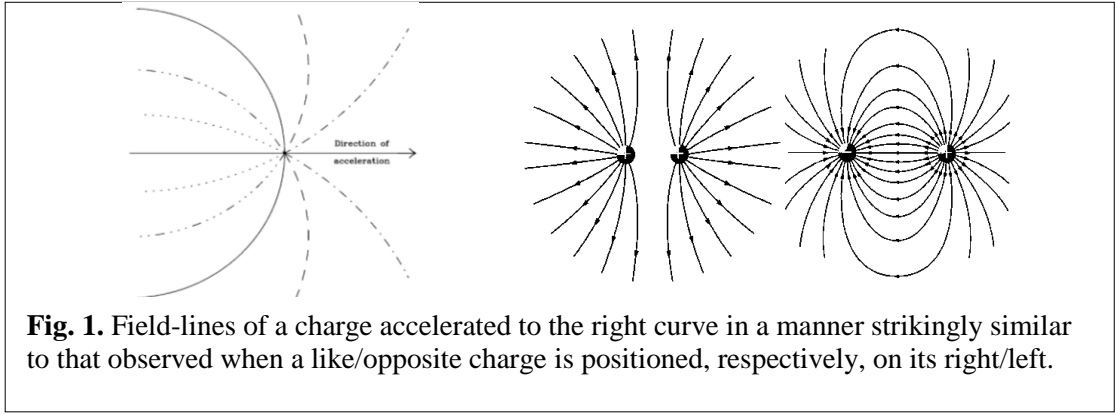

**Fig. 1.** Field-lines of a charge accelerated to the right curve in a manner strikingly similar to that observed when a like/opposite charge is positioned, respectively, on its right/left.

As this field-lines' curvature is strikingly similar to that induced by the presence of a neighboring field (Fig. 1), we proposed [1] a simple symmetric causal relation:

*i)* When a charge is accelerated, its field-lines curve in the opposite direction, storing stress force [2-3] eventually released as electromagnetic radiation. The curved field thereby offers additional inertial resistance to the acceleration, apart from the inertia of the charge's pure mass.

*ii)* When a charge is held at rest in the proximity of an opposite/like charge, its field-lines curve towards/away from the other charge, again storing stress force which is eventually released as kinetic energy.

An appealing symmetry thus emerges: *Field-lines curvature <=> acceleration*. It offers a simple explanation of the electrical attraction/repulsion as the uneven stress forces created within the curved electric field.

Is this a genuine causal symmetry? Our answer in the affirmative was based on an extended review and analysis of the electric fields' mass and stress. Some novel insights derived from this symmetry, bearing on other open questions in electromagnetism, indicate that it is worth pursuing [1].

In this paper, we concentrate on a proposal for a simple experiment which can put the alleged reality of the field to empirical test.

**1. Introducing the Rival Approaches: The Ontological *vs*. Instrumentalist Accounts of the Electric Field**

Field-lines, once introduced to physics by Faraday, are best visualized by his simple aid of test charges' "dust" evenly spread on a surface above the charge. This way, the field of a point charge, a dipole, a charged surface, etc., can be easily detected and even simulated by computer applets. The simulations in Fig. 2, prepared with such an applet [4], are two-dimensional, finite and approximate, hence having some natural limitations, *e.g.*, the charged surface's field is rather that of a charged rod, and the field-lines divergence at its ends make it distinct from an infinite plate/rod. Nevertheless, for illustration purposes they can serve us well: (a) a single point-charge's field, (b) a charged plate's field, and (c) the two fields brought close together.

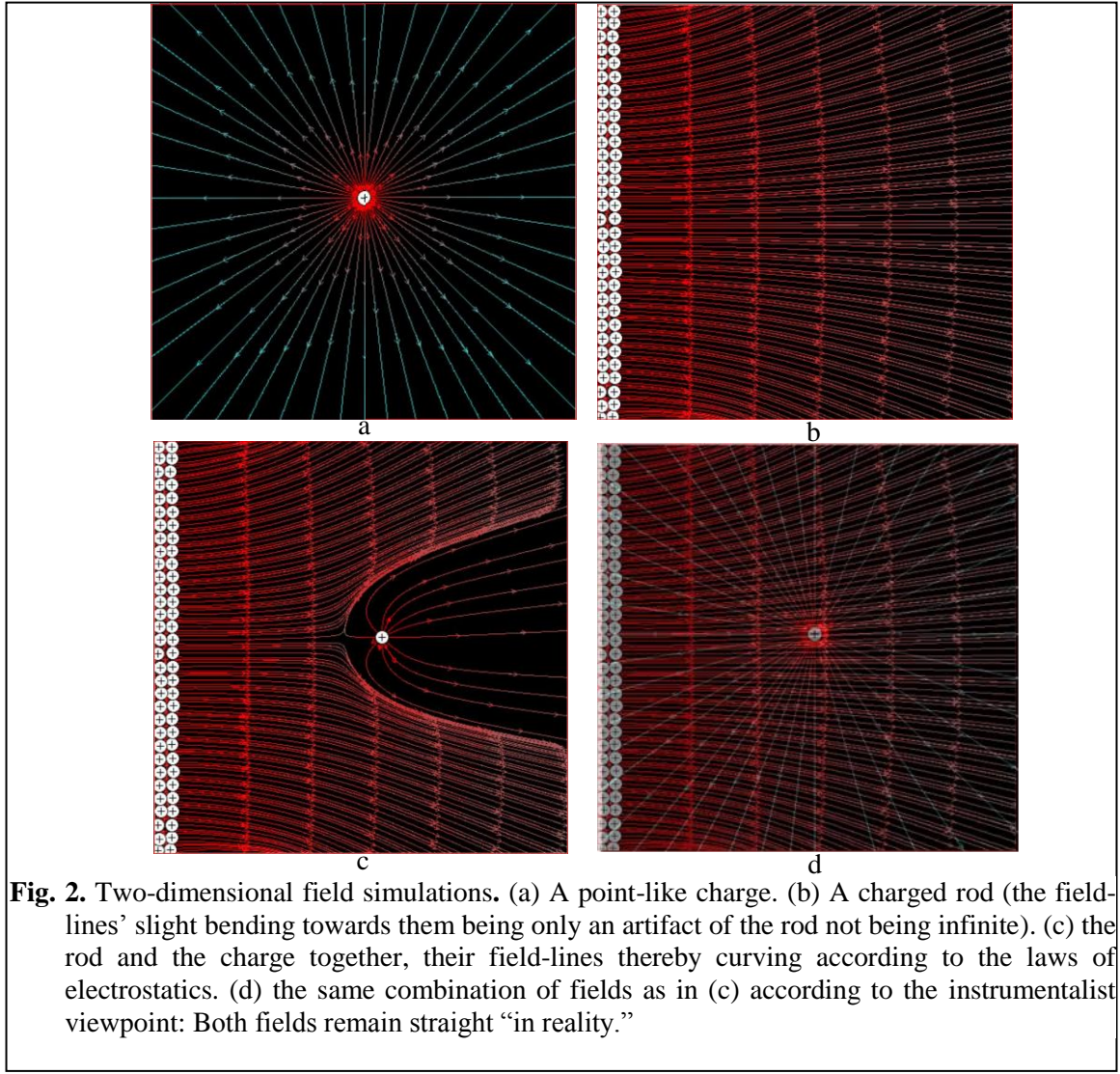

**Fig. 2.** Two-dimensional field simulations**.** (a) A point-like charge. (b) A charged rod (the field-lines' slight bending towards them being only an artifact of the rod not being infinite). (c) the rod and the charge together, their field-lines thereby curving according to the laws of electrostatics. (d) the same combination of fields as in (c) according to the instrumentalist viewpoint: Both fields remain straight "in reality."

This latter case, showing the familiar field-lines curvature, raises this work's main question, as it can be interpreted in two opposing ways:

1. *Ontological*: Fields of different charges interact with one another and then transmit the electric force to one their originators. The field-lines represent real properties of the field, reflecting its mass distribution and tension [5], such that, when the field curves due to some force (Fig. 2c), it stores stress which tends to straighten it back. Hence electric attraction/repulsion.

2. *Instrumentalist*: Field-lines are merely conceptual tools. Fields do not interact with one another, only with charges. Their lines remain straight (Fig. 2d), their apparent bending being an artifact of the fact that in every point in space we only observe the total field, given by its components' superposition [6].

Apparently, both accounts are experimentally equivalent. This is the case in all experiments that test the electric field's electromagnetic interactions: The two approaches predict the same physical result. There is, however, another property of the electric field, seldom attended by experiments, for which the above approaches differ in their predictions, namely, it's *mass*.

Ascribing mass to the electric field goes back to the classical realm. An accelerated charge radiates. Hence, by energy conservation, accelerating it requires greater energy than that needed for a neutral equal mass. It is therefore the field's additional mass that adds inertia to its charge. Even more straightforwardly, the field's mass follows from the relativistic mass-energy equivalence. The electric field squared is proportional to its energy density at a given point in space. Given the entire volume occupied by the charges' electric field (from its classical radius outwards) one can easily calculate the entire energy stored in it. It is then straightforward to calculate the mass associated with the field:

$$m_{field} = \frac{1}{c^2}\frac{1}{8\pi}4\pi\int_{r_q}^{\infty}\frac{q^2}{r^2}dr = -\frac{q^2}{2c^2}\frac{1}{r}\bigg|_{r_q}^{\infty} = \frac{q^2}{2c^2 r_q}. \tag{4}$$

On this property we base our proposed experiment for distinguishing between the above ontological and instrumentalist accounts of the electric field's curvature.

## 2. Deriving Distinct Predictions: The Charged Pendulum Experiment

Consider (Fig. 3a) a small ball with mass *m* hanging with a massless string of length *l* from a frictionless pivot. Let the ball's equilibrium position be the coordinate system's origin point *O*. The gravitational force *mg* acts on it in the *-z* direction.

Let the ball be perpendicularly pushed in the horizontal direction *x*, creating a very small angle $\theta \ll 1$. The ball turns approximately into a mathematical pendulum, swinging along an arc extended the *x-z* plane, such that:

$$x = A_x \sin\omega t, \tag{5}$$

where $\omega = \sqrt{\dfrac{g}{l}}$.

### 2.1. Electrostatics

Next let the (insulating) ball possess also charge *q*. This charge, by Eq. 4, gives the ball some additional (though hardly noticeable) inertia. It will also radiate, eventually slowing down to rest.

In the non-relativistic limit, the equation of motion describing the pendulum motion can be derived by equating the change of the charge's energy with the Larmour power [6]:

$$\frac{dE}{dt} = -P \Rightarrow m\dot{x}\ddot{x} - \frac{mg}{l}x\dot{x} = -\frac{2}{3}\frac{q^2}{c^3}\ddot{x}^2. \tag{6}$$

Evidently, the emitted power *P* is very small, so in order to create an observable effect we should increase the pendulum's charge and/or consider a relativistic motion with a momentarily emitted radiation power of

$$P_{rel} = -\frac{2}{3}\frac{q^2}{c^3}\ddot{x}^2 \gamma^6, \tag{7}$$

where $\gamma = \dfrac{1}{\sqrt{1-(v/c)^2}}$ is the charges' Lorentz factor.

### 2.2. Field-Lines Curvature

Next (Fig. 3b) let us produce the electrostatic state needed to produce the familiar curvature in the charge's field-lines. To the ball's side along the *x-z* plane, parallel to the pendulum's swinging axis, let an insulating plate be positioned, of practically infinite (*i.e.* sufficiently larger than the pendulum's string) size, with a homogeneous equal-sign charge density $2\sigma_0$.

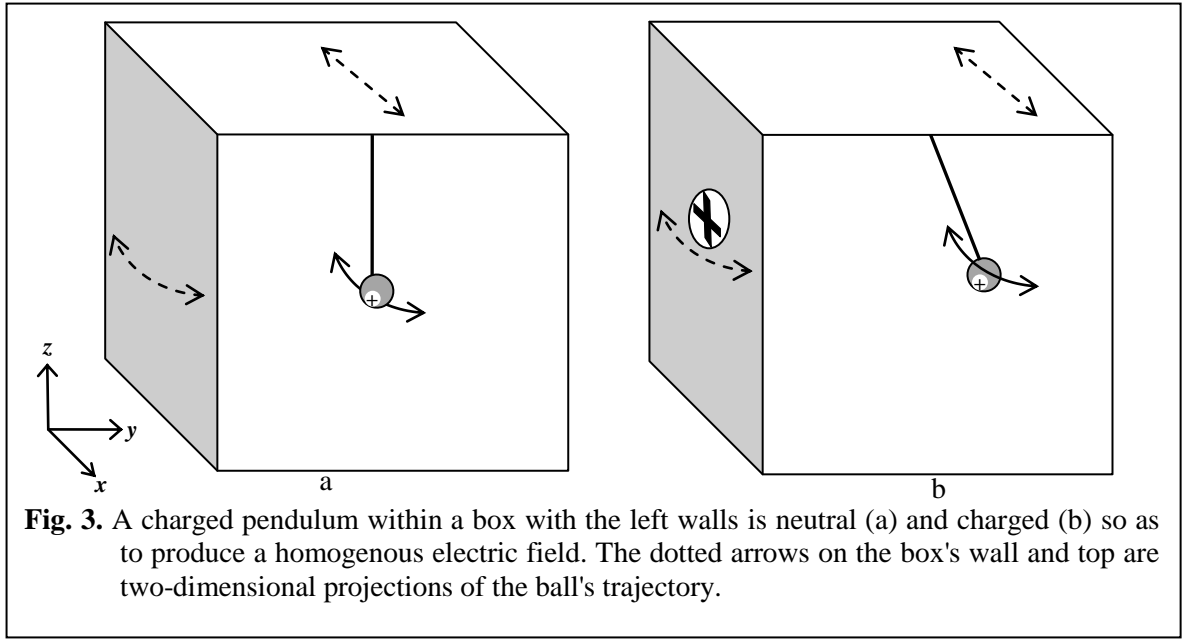

**Fig. 3.** A charged pendulum within a box with the left walls is neutral (a) and charged (b) so as to produce a homogenous electric field. The dotted arrows on the box's wall and top are two-dimensional projections of the ball's trajectory.

The ball, now subject to a force in the $+y$ direction in addition to the $-z$ gravitational force, hangs not straight downwards, but diagonally, displaced (when $\theta$ is small) to $O'$ by

$$y = \frac{4\pi\sigma_0 q l}{mg}, \tag{8}$$

with its string accordingly inclined. When pushed again in the horizontal $x$ direction, the pendulum swings on the tilted $x$-$\tilde{z}$ plane (where $\tilde{z}$ is the tilted z-axis, pointing along the string) with an effective gravitation of: $\tilde{g} = \sqrt{g^2 + (4\pi\sigma_0 q/m)^2}$.

Also, due to electromagnetic energy loss of Eq. 6, the pendulum's frequency will gradually decline,

$$\dot{E} = -P \Rightarrow \omega\dot{\omega} = -\frac{2q^2\gamma^6}{3mc^3}\omega^4 \Rightarrow \dot{\omega} = -\frac{2q^2\gamma^6}{3mc^3}\omega^3 \equiv -k\omega^3$$

$$\Rightarrow \omega(t) = \frac{1}{\sqrt{2}}(kt + \frac{L}{2\tilde{g}})^{-1/2}$$

(9)

In order to minimize radiation damping and simplify the analysis, let us assume henceforth that $\gamma \approx 1$. This, however, should not affect the pendulum's geometrical trajectory. The numerical calculation we perform at the end of Ch.3 justifies these assumptions.

Consider next our two electric fields as visualized by the test particles method (Fig. 4b). They assume the familiar form: The ball's field-lines, otherwise evenly stretched to all directions, now curve to +*y*, away from the plate. Likewise, the plates' field-lines, otherwise parallel and straight in +*y*, now curve in the ball's proximity, bypassing it via *x* and *z*.

### 2.3. Consequences

This is where the above two accounts of the electric field differ in their predictions:

1. *The Instrumentalist Viewpoint* (Fig. 4a): The ball's field remains unchanged, its field-lines curvature merely appearing as an artifact of the superposition principle [6]. As the field's mass distribution is not affected by the electric superposition rule, no displacement of the ball's center of mass should occur.

2. *The Ontological Viewpoint* (Fig. 4b): The ball's electric field-lines curve, affecting the field's mass distribution. Consequently, *the ball's equilibrium point resides in O'', deviating to +$y_0$ from O'*.

Obviously, the two accounts dictate different pendulum motions. With the former, the only change in the pendulum's trajectory, following the equilibrium point's *O-O'* displacement, would be the corresponding inclination of the swinging arc to the *y* direction (Eq. 8.) The ontological account, in contrast, obliges the ball's center of mass (COM) to shift to in *O''*, hence slightly but constantly lag behind its geometric center *O'* with every change of velocity, as it is only the latter that is directly bound by the string. Consequently, the swinging arc should curve not only in the *z* and *x* directions but also, with every swing towards/away from the *O'*, to the +*y* and -*y* direction, respectively.

It is important to note that the magnetic field created by the moving charge is significantly smaller than the electric one so long as v<<c as assumed above. Hence the energy carried by magnetic fields can be made arbitrarily small, hence of no consequence for the ontological prediction.

The trajectory of the pendulum ball with a shifted COM can be approximately calculated by a simple mechanical analogy: Suppose that there is a smaller mass at the ball's +*y* side, connected to it by a spring [7]. The ball's COM now resides opposite to the charged plate. The resulting dynamics simulates that of our shifted COM due to the presumed field curvature.

Our charged pendulum, then, swings like a ball with *inhomogeneous mass density*. As its motion is cyclic, the COM's lag must become increasingly noticeable. Making the ball's and plate's charges greater, while slightly increasing also the ball's mass to preserve the pendulum's axis vertical component, will enhance the effect further.

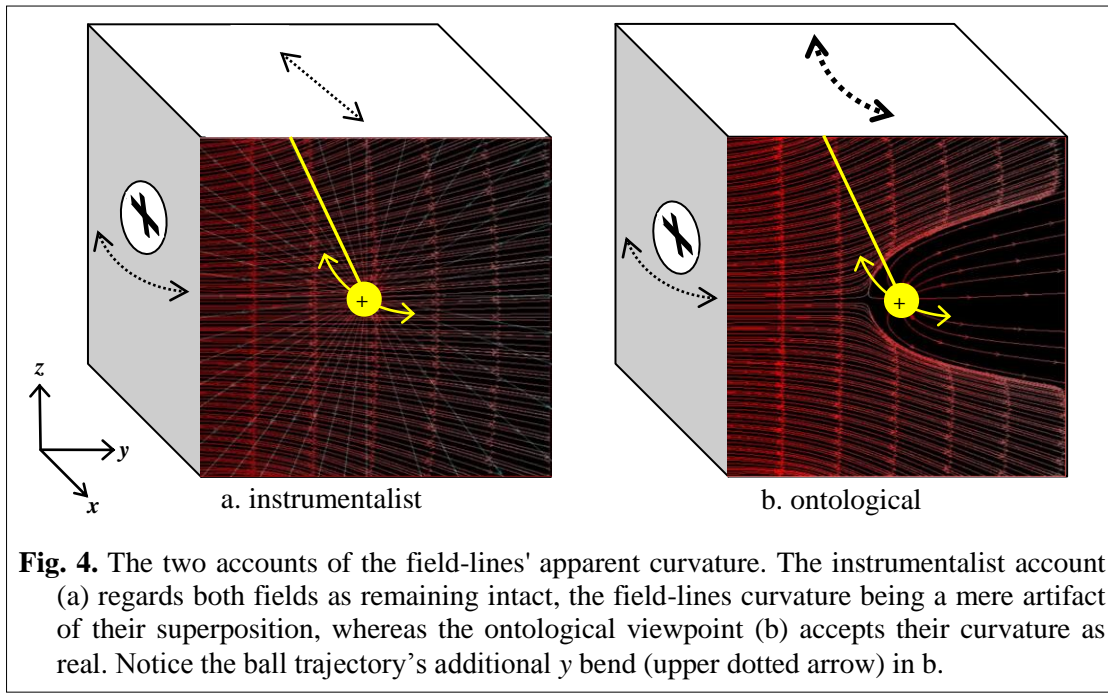

**Fig. 4.** The two accounts of the field-lines' apparent curvature. The instrumentalist account (a) regards both fields as remaining intact, the field-lines curvature being a mere artifact of their superposition, whereas the ontological viewpoint (b) accepts their curvature as real. Notice the ball trajectory's additional *y* bend (upper dotted arrow) in b.

### 3. Locating the Displaced COM

The proposed test for the ontological viewpoint suggested in this paper, *i.e.* the pendulum's additional swing in the *y* direction, is dictated by the amount of displacement of its COM relative to the charge. Let us quantify this displacement.

A few technical issues need however to be addressed first. So far, the experiment has been presented in a simplified, ideal manner. As the plate is ideally supposed to be infinite so as to produce parallel field-lines, its field must extend towards *y* infinitely as well. This infinity will then plague the spatial translation of the ball's COM and consequently the time of the COM's lag behind the charge. To prevent that, a slight modification is needed. Let another, same-size insulating plate, with half the charge density σ0, be placed on the pendulum's other side facing the first plate (Fig. 5). The new configuration's COM will now sift from the charge to a finite distance.

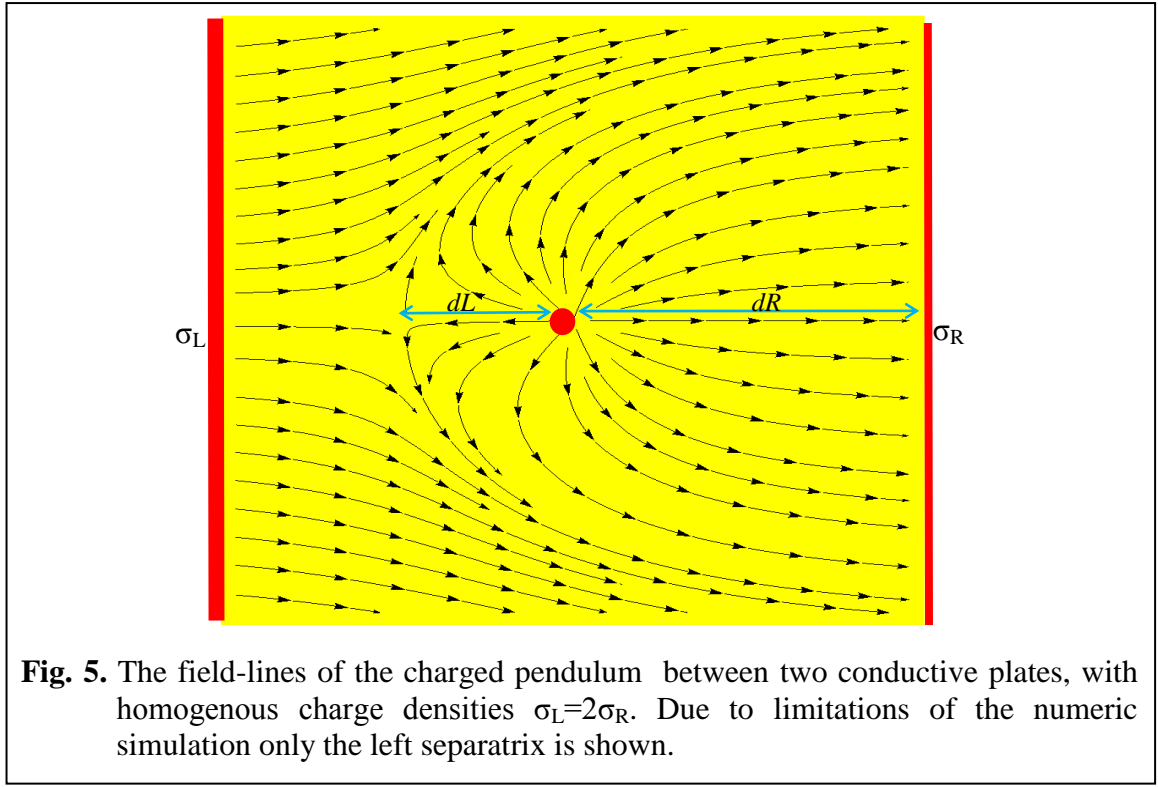

**Fig. 5.** The field-lines of the charged pendulum between two conductive plates, with homogenous charge densities $\sigma_L = 2\sigma_R$. Due to limitations of the numeric simulation only the left separatrix is shown.

The overall resulting field (Fig. 5) can be separated into two regimes. In the origin near the left plate the field appears as that of the left plate alone (charged with $\sigma_L$), except that its lines become slightly sparser due to the contribution of the right plate. Similarly, nearer to the charge the field becomes dominated by the charge itself (*i.e.* all field-lines are those connected to the charge). Between the regimes dominated by the charge and each wall lie two surfaces where the field' direction is parallel to the near wall's surface. This is known as *separatrix* [8-9], marking a boundary between phase curves (such as our field-line) with different properties. It is created when a phase curve meets a hyperbolic fixed point (like the one in which the *y* components of the field are equal). It can be rigorously plotted in space by analyzing Poisson's differential equation. Thus, in the framework of the ontological viewpoint, the two separatrices define the volume of space occupied by the swinging charge's field-lines. An accurate description for the new center of mass of the charge + field is then:

$$Y'_{COM} = \frac{0 \cdot m + \int \frac{E^2 y}{8\pi c^2} dV}{m + m_{field}},\qquad(10)$$

where the charge is taken to be at the origin and the integral on the field is over the entire volume to the right of the separatrix.

Rather than solving the complex mathematical problem involved, we suggest an intuitive lower bound on the COM's displacement . Let us assume an approximately box-like separatrix, of volume

$$V = d_L \times d_L \times (d_L + d_R) \qquad(11)$$

where $L \times L$ is the area of the plates ($L \gg r_q$), $d_R$ the distance between the shifted pendulum and the right plate, and $d_L = \sqrt{\frac{q}{2\pi\sigma_0}}$ the distance between the shifted pendulum and the point along the $y$ axis where its electric field equals the total field of the plates. This box clearly contains the separatrix. Hence, a volume integral on the field mass-density within the box provides a COM somewhat to the left of the analytically calculated COM associated with the field component. Therefore, replacing the integral in the exact equation by the integral confined to this box will result in a lower bound for the COM displacement. The shift of the pendulum's COM thus results from three contributions: the charge's mass, the charge's field and the plates' fields within the separatrix.

Let charged ball's position be again defend as the coordinates' origin, therefore not contributing to the shift. Second, its field's mass $m_{field}$ (by Eq. 4) is not distributed evenly inside the box. However, since, by far, most of the field's energy is concentrated in the charge's vicinity we can neglect this slight asymmetry.

The third contribution comes from the two plates' fields:

$$E_L = 4\pi\sigma_0 \equiv \frac{4\pi Q}{L^2}, \quad E_R = -2\pi\sigma_0 \equiv -\frac{2\pi Q}{L^2}. \tag{12}$$

Their total mass within the separatrix is

$$m_p = \frac{(E_R + E_L)^2}{8\pi c^2} V = \frac{\pi \sigma_0^2}{2c^2} d_L^{\,2}(d_R + d_L). \tag{13}$$

As the plates' field-lines are uniform within the box's volume, their contribution to the new COM is equivalent to that of a point charge with mass (13) located at distance

$$\frac{d_R + d_L}{2} - d_L = \frac{d_R - d_L}{2} \tag{14}$$

to the right of the charge/origin.

Summing up all the contributions,

$$\begin{aligned} Y'_{COM} &\approx \frac{1}{m + m_{field} + m_p}[m \cdot 0 + m_{field} \cdot 0 + m_p \frac{d_R - d_L}{2}] = \\ &= \frac{m_p}{m + m_{field} + m_p} \frac{d_R - d_L}{2} \end{aligned} \tag{15}$$

which is the COM's shifted location.

To quantify this effect we give a numerical estimate based on arbitrarily chosen charge densities. Let the two plates have $L = 100 cm$, $\sigma_0 = 1.2 esu/cm^2$, $2\sigma_0 = 2.4 esu/cm^2$, and the ball $q = 4.8 \cdot 10^{-8} esu$, $m = 10^{-20} gr$, a distance $d_R = 5cm$ from the right plate. Calculating $d_L = 8 \cdot 10^{-5} cm$, $m_p = 7.86 \cdot 10^{-28} gr$, we find $Y'_{COM} = 1.97 \cdot 10^{-7} cm$, resulting in a small correction to the ball's dynamics. Clearly, relativistic corrections are even weaker.

### 4. Considering the Alternatives

The challenge posed at the article's beginning can now be addressed anew: Is it possible to account for the above predicted result without assuming that the two fields really curve so as to change their mass distribution?

For the instrumentalist approach (Sec. 1) to answer this question in the affirmative, it must argue as follows. Although the fields themselves remain unchanged when superposed, their mass distributions *happen to change* just in in the very pattern by which the field-lines "seem to curve."

This *ad hoc* assumption is hardly convincing bearing in mind that, by the superposition rule, the gravitational field-lines of two masses curve *away* from one another (like those of two equal charges), whereas the electric curvature depends on the two charges being like or opposite. In other words, the two electric fields make mass distribution behave differently than two uncharged masses. Therefore, whether the field-lines' curvature is real or only apparent must give different mass distribution.

Another alternative to the ontological account may relate the predicted COM displacement to the radiation-reaction force acting on the charge when non-uniformly accelerating [6]. This force, also known as the Abraham-Lorentz force, is given by

$$\vec{F} = \frac{2}{3}\frac{q^2}{c^3}\dot{\vec{a}} . \qquad (16)$$

However, for the non-relativistic velocities we examined, it is orders of magnitude weaker than the Lorentz force due to its inverse dependence on $c^3$. In addition, it points towards the third time derivative of the position and hence cannot give rise to the constant deviation of the COM.

## 5. Discussion

Classical electrodynamics is commonly believed to have exhausted its foundational questions, leaving further progress to relativity and quantum mechanics. This is far from being the case. Although the electromagnetic field is described by the classical formalism with utmost precision, the field itself, as a physical entity, remains ill-understood. Do field-lines represent objective properties of the field or are they mere mathematical conventions? This question was so far considered merely philosophical. The present case in which the two possible answers give distinct predictions brings the question back to physics.

Moreover, even today, after several decades of advances in relativity and QM, several simple questions concerning the electric field remain as disputed as ever. Does a charge resisting gravity radiate? Why is the advanced solution of Maxwell's equations never observed? Is there self-force between a charge and its own field? Significantly, these and other such fundamental questions were shown to receive fresh twists by the ontological approach [1]. Future work [7] will further explore the consequences of the ontological approach in the contexts of quantum mechanics and general relativity.

## Acknowledgements

It is a pleasure to thank Shay Ben-Moshe, Michael Bialy, Avi Gershon and Shai Kiriati for helpful comments and discussions.

## References

1. A.C. Elitzur., E. Cohen, P. Beniamini, *AIP Conf. Proc*. 1411, pp. 221-244
2. A. Harpaz, *Eur. J. Phys.* **26**, 219-223 (2005).
3. A. Harpaz, *Eur. J. Phys.* 23, 263 (2002).
4. McGuffin, M. J. (2004) Visualization of an electrostatic field. http://www.dgp.toronto.edu/~mjmcguff/research/electrostatic/applet1/main.html.
5. D. R. Rowland, Eur, J. Phys. 28, 201-213 (2007).
6. J. D. Jackson, *Classical Electrodynamics*, New York: Wiley & Sons (1967).
7. D. Grossman, E. Cohen, P. Beniamini, A.C. Elitzur, Forthcoming.
8. Y. Ilyashenko, S. Yakovenko (2008), *Lectures on analytic differential equations*, Graduate Studies in Mathematics, **86**. American Mathematical Society.
9. C. Camacho, P. Sad (1982), *Invariant varieties through singularities of holomorphic vector fields*, Ann. of Math. (2) **115**, no. 3, 579-595.